\documentclass[11pt,psfig,epsfig,graphicx,psfrag]{article}
\usepackage{dcolumn}
\usepackage{amssymb}
\usepackage{lscape}
\usepackage{amsmath}
\usepackage{hyperref}

\newcommand{\be}{\begin{equation}}
\newcommand{\ee}{\end{equation}}

\def\L5{\tilde{\Lambda}}

\def\1{\mathchoice{\rm 1\mskip-4.2mu l}{\rm 1\mskip-4.2mu l}%
{\rm 1\mskip-4.6mu l}{\rm 1\mskip-5.2mu l}}

\newcommand{\beq}{\begin{equation}}

\newcommand{\eeq}{\end{equation}}

\newcommand{\bea}{\begin{eqnarray}}
\newcommand{\eea}{\end{eqnarray}}
\newcommand{\ba}{\begin{eqnarray}}
\newcommand{\ea}{\end{eqnarray}}
\newcommand{\bq}{\begin{quote}}
\newcommand{\eq}{\end{quote}}

\def\a{{\cal C}_1}
\def\b{{\cal C}_2}

\def\m{\mu}
\def\n{\nu}
\def\a{\alpha}
\def\b{\beta}

\title{\bf On horizon structure of bimetric spacetimes}

\begin{document}
\date{}
\maketitle


\begin{center}

 {\large
C\'edric~Deffayet$^{a,}$\footnote{deffayet@iap.fr}
and Ted Jacobson$^{b,}$\footnote{jacobson@umd.edu}}
\\

\vspace{0.5cm}
$^a${\it APC\;\footnote{UMR 7164 (CNRS, Universit\'e Paris 7, CEA, Observatoire de Paris)}, 10 rue Alice Domon et L\'eonie Duquet,\\
 75205 Paris Cedex 13, France}\\
 
 $^b${\it Center for Fundamental Physics, Department of Physics\\
 University of Maryland, College Park, MD 20742-4111 USA}

\vspace{0.5cm}
 \bigskip

{\bf \large Abstract}
\begin{quotation}\noindent

We discuss the structure of horizons in spacetimes with two metrics,
with applications to the Vainshtein mechanism and other examples.
We show, without using the field equations, that if 
the two metrics are static, spherically symmetric, nonsingular, and diagonal 
in a common coordinate system, then a Killing horizon for 
one must also be a Killing horizon for the other. 
We then generalize this result to the axisymmetric case.
We also show that the surface gravities must agree 
if the bifurcation surface in one spacetime
lies smoothly in the interior of the spacetime of the other metric.
These results imply for example that the Vainshtein mechanism 
of nonlinear massive gravity theories cannot work to recover black holes 
if the dynamical metric and the non dynamical flat metric are both diagonal. 
They also explain the global structure of some known solutions of bigravity 
theories with one diagonal and one nondiagonal metric, in which the bifurcation 
surface of the Killing field lies in the interior of one spacetime and on the 
conformal boundary of the other.

\end{quotation}

\end{center}

\section{Introduction}

In various contexts of physics today 
the situation arises where 
two or more Lorentzian metrics may be 
defined on the same spacetime manifold. 
If one of those metrics has a black hole horizon,
or more generally a Killing horizon,
interesting constraints on coordinates and
features of global structure can arise. In particular,
it can be impossible for both metrics
to be diagonal at the horizon, and the
maximal extensions of the two metrics
may not coincide. 
The purpose of this note is to discuss these
issues and give some examples.

Early bimetric theories are those of 
Belinfante-Swihart-Lightman-Lee \cite{BSLL}, 
Isham-Salam-Strathdee \cite{Isham:gm} (``f-g theory"), 
Rosen~\cite{Rosen},   Ni~\cite{Ni}, and  Rastall \cite{Rastall}.
The notion also arises in deformations
of general relativity with a massive graviton,
such as nonlinear 
Pauli-Fierz gravity \cite{ArkaniHamed:2002sp,Damour:2002ws,DRG}. 
Another setting is in theories with a preferred
frame, characterized by a timelike unit vector 
field $u^a$. Then a mode that propagates at 
speed $v_{i}$ relative to the preferred frame 
sees an effective  
metric 
\beq\label{gi}
g^{(i)}_{ab}=g_{ab}+(v_i^2-1)u_au_b.
\eeq
For instance these could be the spin-2,1,0  
modes of Einstein-aether theory~\cite{Jacobson:2000xp, Jacobson:2008aj}, 
or the spin-2 and spin-0 modes of the IR limit
of (completed) Horava gravity~\cite{Horava:2009uw,Sotiriou:2010wn}
or the metrics seen by matter fields that couple
differently to the aether $u^a$. 
A bimetric theory has also recently been introduced for 
MOND \cite{Milgrom:2009gv}. 
Bimetric black holes also arise in the setting of  
Ref. \cite{Dubovsky:2006vk}, where some propagating
field couples to a metric constructed from the ``usual" metric
and a ghost condensate \cite{ArkaniHamed:2003uy}.
Another example arises in the setting of $k$-essence theories 
\cite{Babichev:2007dw}.
Our considerations
apply to all of these settings.

Here we first establish some general properties of bimetric spacetimes
with Killing horizons, 
and then apply and illustrate them with a few specific examples. 
The most important implication is perhaps that regular spherical black holes
can exist in massive gravity theories with a flat nondynamical
metric only if not both of the metrics are diagonal, and then only
if the bifurcation surface is not in the interior of the spacetime common to
both metrics. Some of our results overlap with recent work of Ba\~nados, Gomberoff and 
Pino\cite{Banados:2011hk}, though our techniques differ.

\section{General properties of bimetric Killing horizons}

We consider first static, spherically symmetric 
metrics, and then generalize to the stationary
axisymmetric case.  A spherical coordinate system
$(t,r,\theta,\varphi)$ can always be chosen so that 
any two such metrics locally take the form
\ba
f_{\mu \nu} dx^\mu dx^\nu &=& - J(r) dt^2 + K(r) dr^2 +r^2 d\Omega^2  \label{linef} \\
g_{\mu \nu} dx^\mu dx^\nu &=& -A(r) dt^2 + 2 B(r) dt dr + C(r) dr^2 + D(r)  d\Omega^2 \label{lineg}
\ea
where $d\Omega^2 $ is the metric of the unit 2-sphere,
$d\Omega^2 = d\theta^2 + \sin^2\theta\, d\varphi^2$.
Note that, in general, it is not possible to choose the coordinates
so that both metrics are diagonal. 
Given the two metrics, a number of coordinate independent scalars can be 
constructed using the inverse metrics, for example
$f^{\m\n}g_{\m\n}$, $g^{\m\n}f_{\m\n}$, $f^{\m\n}f^{\a\b}g_{\m\a}g_{\n\b}$, etc. 
(which are all obtained from tracing powers of the matrix 
${\cal M}^\mu_{\hphantom{\nu} \rho} = f^{\mu \nu}g_{\nu \rho}$ \cite{Damour:2002ws}).
Whenever both metrics define regular geometries, all of these invariants
must be regular, even if some components in a singular coordinate system are singular.

We now present three propositions about horizons that hold in this and related situations.
\begin{quote}
{\it Proposition 1}: Suppose the Killing vector $\partial_t$ 
is null at $r=r_H$ with respect to $g_{\m\n}$. Then if both metrics are diagonal and describe 
smooth geometries at $r_H$, $\partial_t$ must also be
null with respect to $f_{\m\n}$ at $r=r_H$. \\

{\it Proof 1a}: The assumptions are that $A(r_H)=0$ and  $B(r) =0$. 
Thus $g^{\m\n}f_{\m\n}= J/A+K/C +  2 r^2/D$, which is
 regular by assumption.  
 In order for both diagonal metrics to have Lorentzian signature, 
 with $t$ translation timelike for $r > r_H$, $J/A$, $K/C$  and 
 $2r^2/D$ must have the same sign, so they cannot cancel. 
But $J/A$ will diverge at $r_H$ 
unless $J(r_H)=0$. 
\end{quote}
A similar observation was made recently
in Ref.~\cite{Banados:2011hk}, 
where it was argued that the presence of the
$g_{\m\n}$ horizon implies that the 1-form $dt$ is singular at $r_H$, 
so that $f_{\m\n}$ will be singular at $r_H$ unless $J(r_H)=0$.  
As discussed below, this reasoning is not conclusive, because at
$r_H$ the 1-form $dr$ becomes proportional to $dt$, so a diverging 
coefficient of $dr^2$ could potentially cancel the singularity
and render the geometry regular.

Next we give an alternative proof that connects
Proposition 1 to properties of the bifurcation surface
of the Killing horizon. This offers a less coordinate
dependent, more global perspective on the 
nature of the proposition, and leads to some generalizations
of the result.
 
We consider the situation where there are two metrics
$f$ and $g$, both  invariant under the flow of the
same Killing vector $\chi$, so that 
${\cal L}_\chi g = 0$ and ${\cal L}_\chi f = 0$.
A Killing horizon is a null hypersurface 
whose null generators coincide with the flow of a Killing 
vector field. Thus, although $f$ and $g$ are assumed to
share the same Killing vector, a Killing horizon for 
$g$ is in general not a Killing horizon for $f$. 

A bifurcation surface of a Killing horizon is a cross-section
of the horizon on which the Killing vector vanishes (and two
sheets of the horizon intersect). For example, 
on the maximal extension of a 
Schwarzschild black hole spacetime,
the spacelike 2-sphere defined by 
the intersection of the future and past horizons
is a bifurcation surface. Another example is
a Rindler horizon $|t|=|z|$ in Minkowski spacetime $\{(t,x,y,z)\}$, 
whose bifurcation surface
is an infinite spacelike plane consisting of the points $(0,0,x,y)$. 

A black hole that forms from collapse does not have a bifurcation
surface in the physical spacetime, but if it ultimately settles
down to a stationary solution with a Killing horizon, then
under rather general conditions it has what might be called 
a ``virtual" bifurcation surface.
Indeed, a  theorem of Racz and Wald  \cite{Racz:1995nh,Racz:1992bp}
proves that if the 
spacetime is static (and therefore with 
a $t$ reflection symmetry), 
or stationary with a $t$-$\phi$ reflection
isometry, and if the 
surface gravity of the horizon is 
nonzero, then
there is an extension of a neighborhood of the
horizon to one 
with a bifurcate Killing horizon. 
The result applies to general 
spacetimes, without assuming
any equations of motion. 
A second result also proved in 
Ref.  \cite{Racz:1995nh} is that any Killing invariant 
tensor field sharing the $t$ or $t$-$\phi$ reflection
symmetry of the metric 
can also be extended globally to the enlarged spacetime.
Now we can give the second proof of Proposition 1:

\begin{quote}
{\it Proof 1b}: If both metrics $f_{\m\n}$ and $g_{\m\n}$
are diagonal then $g_{\m\n}$ shares the
$t$ reflection symmetry  of $f_{\m\n}$.
If the surface gravity of the $g$-horizon is nonzero, then
the Racz-Wald theorem implies that
both metrics can be extended to a
regular bifurcation surface of the $\partial_t$ Killing 
horizon for $g$. The scalar $f_{\m\n}\chi^\m\chi^\n = J(r)$ 
vanishes at the bifurcation surface where $\chi^\m=0$,
and it cannot change along the Killing flow, so it vanishes
everywhere at $r=r_H$.
\eq

This method of reasoning extends to the stationary
case, in which spherical symmetry need not be assumed.

\bq
{\it Proposition 2}: Suppose  that $\chi^\m$ is 
a Killing vector for metrics
$g_{\m\n}$ and $f_{\m\n}$, and that
$g_{\m\n}$ has a $\chi^\m$  Killing horizon 
with nonvanishing
surface gravity. Suppose further that both geometries
are regular and possess the $t$-$\phi$ reflection
isometry. Then the horizon is also a horizon for 
$f_{\m\n}$.\\

{\it Proof 2}: By the Racz-Wald theorem,
$f_{\m\n}\chi^\m\chi^\n$ must smoothly extend to
the birfurcation surface, where it must vanish.
However, it is constant along the Killing flow,
so must vanish everywhere on the
$g_{\m\n}$-Killing horizon, which is therefore also
a Killing horizon for $f_{\m\n}$.
\eq

The results discussed so far do not preclude the 
coexistence of two geometries,
one with a Killing horizon
and one without.
Rather they only imply that the nonhorizon 
geometry cannot possess the $t$-$\phi$ reflection symmetry. 
For example, the presence of a nonzero $t$-$r$ 
component in the metric (\ref{lineg}) can
allow both geometries
to be regular at the horizon, ({\it cf}.\ Sec. \ref{SchwFlat} for more details.) 
In this example,  both metrics are static and hence possess a
$t$ reflection symmetry, but that symmetry is not the same 
for $g_{\m\n}$ as it is for $f_{\m\n}$, since
the Killing field $\partial_t$ is $f$-orthogonal but
not $g$-orthogonal
to the constant $t$ surfaces. 

When both geometries are regular and
coexist at a Killing horizon
of one of the metrics that is not a Killing horizon
for the other metric, there is a
global consequence of the mismatch. Namely,
the bifurcation surface of the $g$ spacetime  
cannot lie in the 
interior of the $f$ spacetime.
This follows by the same argument used in Proof 2. 
In fact, this result can be strengthened further to require 
matching surface gravities.

\bq
{\it Proposition 3}: If a Killing horizon of a metric
$g$ has a bifurcation surface that lies in the 
interior of the spacetime of another metric $f$ with the 
same Killing vector, then it must also be 
a Killing horizon of $f$, and with the same surface gravity.\\

{\it Proof 3}: The necessity that the horizon is also
an $f$ Killing horizon is clear from Proof 2. 
The equality of surface gravities
follows from the fact that the surface gravity $\kappa$
can be defined in an entirely metric-independent
way at the bifurcation surface 
(cf.\ section III of \cite{Jacobson:1993vj}):

\bea
2\kappa^2 &=& -(\nabla_\a\chi_\b)(\nabla^\a\chi^\b)\\
&=& (\nabla_\a\chi^\b)(\nabla_\b\chi^\a)\\
&=& (\partial_\a\chi^\b)(\partial_\b\chi^\a).\label{dchidchi}
\eea
In the second line we used Killing's equation 
$\nabla_{(\a}\chi_{\b)}=0$, and in the last line
we replaced the covariant derivative by the
partial derivative in any coordinate system, 
since the terms with Christoffel symbols
vanish because $\chi^\a=0$ at the bifurcation surface. 
(Note that, of course, Eq. (\ref{dchidchi}) 
can be used only in a coordinate
system that is regular at the bifurcation surface.
The argument requires only the existence of such 
a coordinate system, which imposes no constraint 
once it is assumed that both geometries are regular there.) 
In the spherically symmetric case, the result of proposition 3 
was proved in Ref.~\cite{Banados:2011hk} using coordinate-based arguments.

\eq

\section{Applications and examples}
The propositions proven in the previous section have various 
interesting applications. We now give some examples, starting 
with massive gravity.

\subsection{Massive gravity and the Vainshtein mecanism}
A long-standing question concerning ``massive gravity" 
is whether the so-called van-Dam-Veltman-Zakharov 
discontinuity \cite{vanDam:1970vg} can be avoided. 
The ``discontinuity" is the fact 
that a massive graviton, as defined by the 
only ghost free quadratic action, the Pauli-Fierz action \cite{Fierz:1939ix}, 
leads to physical predictions, such as light bending, which are 
significantly different from those of linearized general relativity, 
however small the graviton mass may be. A way to recover GR, 
proposed by A. Vainshtein~\cite{Vainshtein:1972sx}, is to 
properly take into account nonlinearities of the theory.
A simple nonlinear theory of massive gravity is obtained by considering 
a bimetric theory with one dynamical metric $g_{\mu\nu}$ and one 
nondynamical metric $f_{\mu \nu}$ which is chosen to be flat. 
A suitable coupling between the two metrics is chosen such that the 
theory, when linearized around flat spacetime, matches the Pauli-Fierz 
theory (in the terminology of \cite{Damour:2002ws}, such a theory is in 
the Pauli-Fierz universality class). 

Various investigations of the Vainshtein proposal have been made in 
such a theory, looking for static and spherically symmetric solutions 
with a matter source \cite{Damour:2002gp,DefV}. Should the Vainshtein 
 mechanism be valid, it would be expected that one can recover, 
for some range of distances below a so-called Vainshtein radius, 
a solution for the dynamical metric very close to the Schwarzchild 
solution outside the source. Following Vainshtein's original proposal 
\cite{Vainshtein:1972sx}, this is expected to occur when both metrics 
take the form  (\ref{linef},\ref{lineg}), and when the nondiagonal coefficient 
$B(r)$ is chosen to vanish. This was in fact shown recently to occur for low 
density sources \cite{DefV}\footnote{Note also that a previous reference 
\cite{Damour:2002gp} reported to have failed to find such solutions.}, 
but nothing is known for dense sources or black holes. 

The results of the previous section allow us to settle this issue for black holes. 
Indeed proposition 1 shows that, provided the dynamical metric $g_{\mu \nu}$ 
has a Killing horizon and both metrics are diagonal, the nondynamical metric 
must also have a Killing horizon at the horizon of the black hole. This cannot be 
the case, however, since the nondynamical metric is considered flat, so has only
planar Killing horizons, not spherical ones. 
This shows that the Vainshtein mecanism, as it is usually formulated (i.e.\ with 
both metrics diagonal), cannot work to recover black holes. It also raises interesting 
questions on how the transition occurs from the low density nonsingular 
solutions found in Ref. \cite{DefV} to the mandatory singular behavior for black holes.

Our results can also shed some light on the solutions of a family of theories of massive gravity which was recently proposed \cite{DRG} as a candidate to avoid well known pathologies of standard nonlinear Pauli-Fierz theories \cite{BDeser}.  Following \cite{Gruzinov}, we will refer to this family of theories as PF2. PF2 theories are similar to the kind of massive gravities considered in this subsection, with one dynamical metric $g_{\mu \nu}$ and one nondynamical flat metric $f_{\mu \nu}$ (note that the flat metric is often written in a so-called nonunitary gauge, thanks to four scalar fields, where it does not have the canonical Minkowksi form).
Static spherically symmetric solutions for FP2 theories with both metrics diagonal have recently been worked out \cite{Nieu,Gruzinov} (see also \cite{Koyama,deRham:2011pt}). Some of the solutions presented in Refs.~\cite{Nieu,Gruzinov} possess a (generally singular) ``black hole horizon" and are nonsingular everywhere outside the horizon. For a subset of those solutions (obtained for a subset of the PF2 theories) the $g$ 
geometry is nonsingular on the horizon as well. However, our results show that there can exist no nonsingular extension of the nondynamical geometry 
to the horizon (and beyond) such that it stays flat (as required by the way the theory is defined). 
It follows that all  the solutions presented in Refs.~\cite{Nieu,Gruzinov} cease to exist at the horizon. 

\subsection{Examples with Schwarzschild and flat spacetime}
\label{SchwFlat}
To illustrate
the previous point, notice that close to the horizon the geometry of those solutions is similar to the one obtained from a Schwarzschild spacetime $-q dt^2 + q^{-1} dr^2 + r^2 d\Omega^2$,
with $q=1-2M/r$, 
together with the flat metric $-dt^2 + dr^2 + r^2 d\Omega^2$ defined using the same
coordinates. Outside the horizon both metrics
are regular, however the propositions proved here show that 
there exists no regular extension of this flat 
geometry to a neighborhood that includes any part of the horizon. 

A contrasting example is provided by the Schwarzschild metric expressed using ingoing
Eddington-Finkelstein coordinates, $-q dv^2 + 2 dv dr$, together with the 
flat metric $-dv^2 + 2 dv dr$, where now the angular part of the metric is implicit.
Clearly both of these metrics are regular in the region outside and on the horizon, and 
inside the black hole region. However, while the black hole geometry is incomplete
in this coordinate patch (and can be extended to the full Kruskal-Szekeres geometry),
the flat metric in these coordinates is complete as it stands, and does not 
include the bifurcation surface which lies at $v=-\infty$. This illustrates the
general fact that the bifurcation surface cannot lie in the interior
of the spacetime that does not have a horizon at the same location~\cite{Blas:2005yk}.

The role of the off-diagonal terms in the flat metric can be appreciated by 
expressing this last example using the Schwarschild time coordinate, with 
$dt=dv-q^{-1}dr$. The black hole line element then takes the usual
diagonal Schwarzschild form, while the flat metric is 
 $-dt^2 +2(1 - q^{-1})dt dr +(2q^{-1}-q^{-2})dr^2$. 
Since this arose by a coordinate
transformation from the previous example, 
we know that this flat geometry can be extended across 
the future horizon. On the other hand, it remains
true that the 1-form $dt$ is singular and null on the horizon. The last 
expression defines a regular metric because the 1-form $dr$ is
also null, hence proportional to $dt$ on the horizon, and the divergent
factors of $q^{-1}$ are just what is needed to allow the
$dt dr$ and $dr^2$ terms together to cancel the divergence
of $dt^2$.

\subsection{Bigravity black holes}
Some regular solutions with black holes are known in theories with two metrics. 
This is the case \cite{BHbig,Damour:2002gp} in the ``f-g theory" (or ``strong gravity") 
formulated in the 70's by Isham, Salam and Strathdee in the context of strong 
interactions \cite{Isham:gm}, and recently  revived
to deal with cosmic acceleration \cite{Damour:2002wu}.  
Those solutions (see also \cite{Come}) are in the 
nondiagonal form (\ref{linef},\ref{lineg}) (i.e. with $B$ nonzero), with 
both metrics  in the Schwarzschild-(anti)-de-Sitter family of solutions of 
GR (the two metrics can have different (anti)-de-Sitter radii as well as Schwarzschild radii).  
Hence, when considered individually, both metrics can be extented to spacetimes 
(the maximal extensions of Schwarzschild-(anti)-de-Sitter spacetimes) containing
bifurcation spheres. However, the bifurcation sphere of one metric always lies outside 
(namely at infinities) of the patch where the other metric is defined \cite{Blas:2005yk}. 

The necessity of this  follows from Proposition 3 of the previous section. Indeed, 
when the horizons of the two metrics do not coincide---which is usually the case in 
the solutions considered, given that the (anti)-de-Sitter radii as well as 
Schwarzschild radii appear as integration constants and can take different 
values for the two metrics---the same is true for the bifurcation surfaces. 
Hence the bifurcation surface of one metric cannot lie in the interior of the 
patch covered by the other.

A similar situation occurs in the black hole
solutions discussed in 
Ref. \cite{Dubovsky:2006vk}, where one metric is the ``usual" one and the 
other is the one seen by some propagating field coupled to a ghost condensate 
\cite{ArkaniHamed:2003uy}, but both metrics are in fact in the standard 
Schwarzschild family. More generally, it occurs when considering two of the metrics of
the type (\ref{gi}) with different values of $v_i$, when $g_{ab}$ and $u^a$ are
both invariant under some Killing flow. For instance, it occurs also in the
Einstein-aether/Horava black hole solutions of Refs.\cite{Eling:2006ec,Barausse:2011pu}.

As a last example, we discuss the case of 
two Schwarz\-schild black holes with a common Killing horizon, 
but different surface gravities. 
According to proposition 3, the bifurcation surfaces must lie outside 
of the patch where both geometries 
are simultaneously non singular. 
For example, consider a
Schwarzschild metric $f = -q dv^2 + 2 dv dr + r^2 d\Omega^2$
of mass $M$
expressed using ingoing Eddington-Finkelstein coordinates,
and using the same coordinates a second metric 
$g= -q dv^2 + dv dr + r^2 d\Omega^2/4$. The second metric $g$ 
takes the standard ingoing-Eddington-Finkelstein form of the Schwarschild geometry
with mass $M/2$ in terms of the new radial coordinate $r'=r/2$. The metrics $f$ and $g$ 
share a common Killing horizon at $r=2M$, however
the surface gravity of $g$ is twice as large as that of $f$. 
This $(v,r)$ coordinate patch defines the maximum region wherein both
geometries are simultaneously regular. To see that, note that
the coordinate change $dt = dv-q^{-1}dr$, puts $f$ in the standard diagonal Schwarzschild form,
and gives $g$ the form $-q dt^2 - dt dr + r^2 d\Omega^2/4$. The coordinates $(t,r')$ are 
thus {\it outgoing} Eddington Finkelstein coordinates for $g$. These coordinates
are regular on the past horizon of $g$ on either side of the bifurcation surface, but the components
of $f$ are singular there, so the two metrics cannot both be extended
across this horizon.
Following constant $t$ curves in the $f$ and $g$ spacetimes, 
one can see that the bifurcation surface of $f$ 
is spread over the past horizon of $g$. 
Taking $t$ to $-\infty$ shows that the part of the past horizon of
$f$ to the past of the bifurcation surface corresponds to past 
timelike infinity ($i^-$) of the $g$ spacetime. (The 
part to the future of the bifurcation surface corresponds to future
timelike infinity ($i^+$) of the other asymptotic region of the $g$ spacetime.)
In this case, in contrast to the previously mentioned examples, 
the bifurcation surface of $f$
does not meet the 
$g$ spacetime at the conformal boundary, but rather 
at interior points. However, it is not smoothly embedded in the $g$
spacetime, so it can have a different surface gravity. 

Analytic continuation of the time coordinate yields
a simple euclidean analog of this example.
Consider $dr^2 + r^2 d\theta^2$ and $dr^2 + 4r^2 d\theta^2$.
The analog of the bifurcation surface is the origin
$r=0$.  If we fix the period of $\theta$ at $2\pi$, the
first metric describes a flat plane, whereas if the period is $\pi$ the 
second metric describes a flat plane. The different periods correspond to
different surface gravities in the Lorentzian continuation. 
The point $r=0$ of each geometry 
is in the interior of the other flat plane,  
but each geometry has a conical singularity on the manifold of the other
one. Thus no neighborhood of that point exists for which both 
geometries are regular.


\section*{Acknowledgments}
The authors thank the organizers of Peyresq Physics XIII where this work was initiated. 
This work was partially supported by
NSF grants PHY-0601800 and PHY-0903572.

\end{document}